\documentclass{article} 
\usepackage{iclr2025_conference,times}
\usepackage{pifont}

\usepackage{amsmath,amsfonts,bm}

\def\eqref#1{equation~\ref{#1}}

\def\1{\bm{1}}

\DeclareMathAlphabet{\mathsfit}{\encodingdefault}{\sfdefault}{m}{sl}
\SetMathAlphabet{\mathsfit}{bold}{\encodingdefault}{\sfdefault}{bx}{n}

\usepackage[hyperfootnotes=false]{hyperref}
\usepackage{url}
\iclrfinalcopy

\usepackage[hyperfootnotes=false]{hyperref}
\usepackage{url}
\usepackage{multirow}
\usepackage{svg}

\usepackage{xcolor}

\definecolor{lightpink}{HTML}{ed9782}
\definecolor{lightblue}{HTML}{5395f5}
\definecolor{lightgreen}{HTML}{efd08f}
\definecolor{grey}{HTML}{b3b3b3}

\usepackage[most]{tcolorbox}

\usepackage{algorithm}

\usepackage{enumitem}
\usepackage{tabularx}

\usepackage{makecell}

\usepackage{tabularx}
\usepackage{multirow}

\usepackage[normalem]{ulem}
\useunder{\uline}{\ul}{}

\usepackage{times}
\usepackage{moreverb}
\usepackage{graphicx}
\usepackage{mathrsfs}
\usepackage{bm}
\usepackage{url}
\usepackage{amsmath}
\usepackage{amssymb}

\usepackage{tikz}
\usepackage{array}
\usepackage{times}
\usepackage{helvet}
\usepackage{courier}
\usepackage{multirow}
\usepackage{mathrsfs}
\usepackage{graphicx}
\usepackage{enumitem}
\usepackage{graphicx}
\usepackage{blindtext}
\usepackage{algorithm}
\usepackage{lineno,hyperref}
\usepackage[marginal]{footmisc}
\usepackage[utf8]{inputenc}
\usepackage[english]{babel}
\usepackage{epstopdf}
\usepackage{array}
\usepackage{multirow}
\usepackage{booktabs}
\usepackage{amsmath}
\usepackage{wrapfig}
\usepackage{times}
\usepackage{latexsym}

\usepackage{xcolor}

\usepackage{amsmath}
\usepackage{amssymb}
\usepackage{algorithm}
\usepackage{algpseudocode}
\usepackage{multicol}

\usepackage[T1]{fontenc}

\usepackage[utf8]{inputenc}

\usepackage{microtype}

\usepackage{inconsolata}

\usepackage{graphicx}
\usepackage[normalem]{ulem}
\useunder{\uline}{\ul}{}
\usepackage{subfigure}

\title{Geneshift: Impact of different scenario shift on Jailbreaking LLM}

\author{
  \textbf{Tianyi Wu\textsuperscript{1,2,3}\thanks{Equal Contribution}}\quad
  \textbf{Zhiwei Xue\textsuperscript{1,3}\footnotemark[1]}\quad
  \textbf{Yue Liu\textsuperscript{1,2,3}\footnotemark[1]}\quad
  \textbf{Jiaheng Zhang\textsuperscript{3}}\quad
  \textbf{Bryan Hooi\textsuperscript{2,3}}\quad
  \textbf{See-Kiong Ng\textsuperscript{2,3}}\\[6pt]
  \textsuperscript{1}Integrative Sciences and Engineering Programme, NUS Graduate School, National University of Singapore \\
  \textsuperscript{2}Institute of Data Science (IDS), National University of Singapore \\
  \textsuperscript{3}Department of Computer Science, School of Computing, National University of Singapore \\
  \small{\texttt{tianyi\_wu@u.nus.edu}\quad \texttt{zhiweixue@u.nus.edu}\quad \texttt{yliu@u.nus.edu}}
}

\begin{document}

\maketitle

\begin{abstract}

Jailbreak attacks, which aim to cause LLMs to perform unrestricted behaviors, have become a critical and challenging direction in AI safety. Despite achieving the promising attack success rate using dictionary-based evaluation, existing jailbreak attack methods fail to output detailed contents to satisfy the harmful request, leading to poor performance on GPT-based evaluation. 
To this end, we propose a black-box jailbreak attack termed \textbf{GeneShift}, by using a genetic algorithm to optimize the scenario shifts. Firstly, we observe that the malicious queries perform optimally under different scenario shifts. Based on it, we develop a genetic algorithm to evolve and select the hybrid of scenario shifts. It guides our method to elicit detailed and actionable harmful responses while keeping the seemingly benign facade, improving stealthiness. Extensive experiments demonstrate the superiority of GeneShift. Notably, GeneShift increases the jailbreak success rate from 0\% to 60\% when direct prompting alone would fail. We will open-source our code.



\begin{center}
\color{red}Warning: this paper contains potentially harmful text.
\end{center}

\end{abstract}

\section{Introduction}
\label{sec:intro}
Large Language Models (LLMs) \citep{achiam2023gpt,dubey2024llama} are indispensable for tasks like knowledge-seeking, content generation, and planning, yet they remain vulnerable to attacks that bypass safety mechanisms and elicit harmful responses. Although white-box methods (e.g., GCG \citep{GCG}) show promising success rates, they require model weights and intensive optimization, limiting practicality for closed-source LLMs. In contrast, black-box methods like PAIR \citep{PAIR}, ArtPrompt \citep{Artprompt}, and SelfCipher \citep{SelfCipher} circumvent these constraints using iterative prompts, artistic language, or cipher techniques. In our experiments, we observe that recent state-of-the-art black-box models indeed achieve good performance under dictionary-based evaluation, which detects the success of the attack by simply checking whether the response contains refusal keywords (a list of <40 keywords). However, with our case studies and further experiments, we find that these methods fail due to either producing insufficiently detailed harmful responses or non-harmful responses that do not contain refusal keywords, thus being misclassified by dictionary-based metrics.

To address this shortfall, we introduce \textbf{GeneShift}, a novel black-box attack that employs a genetic algorithm to optimize scenario shifts. We find that different malicious prompts perform optimally under varying scenario shifts. By evolving and selectively blending these shifts, GeneShift elicits more detailed and actionable harmful responses while retaining a benign facade. Experiments on GPT-4o mini confirm GeneShift’s effectiveness, demonstrating a marked improvement in attack success rates.

Our main contributions are as follows: \begin{itemize} 
\item We show that aligning distinct scenario shifts with specific malicious behaviors yields the most detailed harmful responses, emphasizing the need for tailored scenario elements. 
\item We demonstrate that malicious prompt efficacy is dependent on contextual configuration, with optimal performance emerging only under the right scenario shifts.
\item We propose GeneShift, a black-box jailbreak attack that employs a genetic algorithm to optimize scenario shifts, enabling malicious prompts to bypass LLM guardrails while appearing benign.
\item Our experiments on GPT-4o mini reveal that GeneShift boosts the jailbreak success rate from below 1\% to 60\%. 
\end{itemize}

\section{GeneShift}
\label{sec:method}

\subsection{Scenario Shift for Enhanced Detail} Direct requests for harmful content often trigger immediate refusal from LLMs. To circumvent this, attackers often embed the malicious query in a broader, seemingly benign context--a \emph{Scenario Shift (CS)}--prompting more elaborate responses that can still fulfill harmful intentions. However, a single, fixed scenario may fail if it does not sufficiently align with the malicious query or misleading enough to bypass the model’s guardrails, motivating an automated approach to systematically discover more suitable transformations.

\subsection{GeneShift: A Genetic Algorithm for Scenario Optimization} To automate the search for optimal scenario shifts, we propose a black-box jailbreak framework called \textbf{GeneShift}. It leverages a genetic algorithm (GA) to explore, evaluate, and refine various transformations, ultimately generating a single-turn prompt that elicits the desired harmful response.
\\[3pt] 
\textbf{Gene Design.}  \citet{yu2024dontlistenmeunderstanding} studied existing jailbreak attacks and classified them into ten categories, we adopt their transformation categories and introduce an additional transformation rule, forming a gene database \( \mathcal{G} := \{ \tau_1, \tau_2, \dots, \tau_M \} \).
\\[3pt] 
\textbf{Population Initialization.}  
We initialize a population \( \mathcal{P} = \{(p_j, g_j)\}_{j=1}^N \), where each candidate's gene \( g_j \) is generated by randomly selecting a subset of \( z_j \) number of distinct transformation rules from the gene database \( \mathcal{G} \). The number of selected transformation rules, i.e., \( z_j \), is sampled from a uniform distribution $z_j \sim \mathcal{U}(1, Z)$, where$\quad Z = 4$.
Each candidate's gene \( g_j \subset \mathcal{G} \) consists of a set of distinct, randomly selected transformation rules, i.e., \( g_j = \{ \tau_{i_1}, \tau_{i_2}, \dots, \tau_{i_{z_j}} \} \), where \( \tau_{i_k} \in \mathcal{G} \) and \( i_1, i_2, \dots, i_{z_j} \) are distinct indices. The corresponding candidate jailbreak prompt \( p_j \) is then generated by the LLM using the selected transformation rules:
\\[-5pt] 
\begin{equation}
    p_j = \text{LLM}(g_j), \quad g_j \subset \mathcal{G}, \quad |g_j| = z_j
\label{Equation:population_init}
\end{equation}

Each pair \( (g_j, p_j) \) is stored to track the transformations applied in future iterations.
\\[3pt] 
\textbf{Fitness Evaluation.}  
To evaluate the fitness of each candidate jailbreak prompt \( p_j \), we utilize a two-step process involving both the attack model and a judge LLM. Given a response \( r_j \) from the attack model, we instruct the judge LLM to classify \( r_j \) into one of six quality categories. The detailed design of the score can be found in Appendix. 

We define the fitness function \( F(p_j) \) of the candidate prompt \( p_j \) as the score assigned by the judge LLM:
\\[-5pt] 
\begin{equation}
    F(p_j) = \text{score}(r_j), \quad r_j = \text{AttackModel}(p_j)
\end{equation}

The fitness \( F(p_j) \) is sampled from a discrete probability distribution \( P(F(p_j) = k) \), where \( k \in \{1, 2, \dots, 6\} \), reflecting the likelihood of each response category based on the attack model's behavior.
\\[3pt] 
\textbf{Crossover.}  
Before performing crossover, we preserve the top \( k \) performing candidates as elites \( \mathcal{E} \subset \mathcal{P} \) to ensure that high-quality individuals are carried over to the next generation. For the remaining candidates, we select parents based on fitness-proportional selection, where the probability of selecting a parent \( p_j \) is:
\begin{equation}
    P(p_j) = \frac{F(p_j)}{\sum_{i=1}^{N} F(p_i)}, \quad j = 1, \dots, N.
\end{equation}

Given two parents \((p_a, g_a)\) and \((p_b, g_b)\), the offspring’s gene \( g_{\text{child}} \) is created by randomly swapping 1 or 2 transformation rules between the two parents. Let \( I \in \{0, 1\}^{z_j} \) be a binary mask indicating which genes are swapped between the parents. The offspring’s gene is then represented as:
\\[-10pt] 
\begin{equation}
    g_{\text{child}, i} = I_i g_{a, i} + (1 - I_i) g_{b, i}, \quad I_i \sim \text{U}\{0, 1\}.
\end{equation}
Thus, each gene \( g_{\text{child}, i} \) is inherited either from parent \( g_a \) or \( g_b \), depending on the value of \( I_i \). The offspring prompt \( p_{\text{child}} \) is then generated by passing the crossovered gene to the LLM similar to Equation \ref{Equation:population_init}:
$p_{\text{child}} = \text{LLM}(g_{\text{child}})$.
\\[3pt] 
\textbf{Mutation.}  Mutation introduces further diversity into the population by modifying or expanding some of the genes in the offspring. For each gene \( g_{\text{child}, i} \), mutation occurs with a probability \( p_{\text{mut}} \). If mutation occurs, the operation is randomly chosen with equal probability: (1) Switch, where the gene is replaced by a randomly selected gene from the gene pool \( \mathcal{G} \); or (2) Add, where a new random gene is appended to the gene sequence:
\begin{equation*}
    g_{\text{child}, i} =
    \begin{cases}
    g_{\text{rand}} & \text{if } O < 0.5 \quad  \\[5pt]
    g_{\text{child}, i} \cup \{g_{\text{rand}}\} & \text{if } O \geq 0.5 \quad 
    \end{cases}
\end{equation*}
\begin{equation}
    I_i \sim \text{Bernoulli}(p_{\text{mut}}), \quad O \sim \text{Uniform}(0, 1), \quad g_{\text{rand}} \sim \mathcal{G}.
\end{equation}
This mutation process ensures new genetic material is introduced and maintains population diversity across generations. The switch operation directly replaces existing genes, while the add operation introduces additional genes, enriching the offspring’s potential solution space.
\\[3pt] 
\textbf{Termination Criteria.}  
The genetic algorithm terminates when one of the following conditions is satisfied:
1) The maximum number of iterations \( T \) has been reached.
2) The number of candidates with fitness scores \( F(p_j) > 5 \) meets or exceeds a predefined threshold \( \theta \), expressed as: $\sum_{j=1}^{N} \mathbb{I}(F(p_j) > 5) \geq \theta$, where \( \mathbb{I}(\cdot) \) is the indicator function. The best-performing candidates is returned as the output of the algorithm.

\section{Experiment}

\subsection{Evaluation Metrics}

\paragraph{Dictionary-based Evaluation.}
Following \citet{GCG}, we check whether the model’s response contains any predefined refusal phrases (see Table~\ref{tab:rejection_dictionary}). If such phrases appear, the attack is considered a failure; otherwise, it is considered successful under this metric, denoted as \(\text{ASR-DICT}\).

\paragraph{GPT-based Evaluation.}
As dictionary-based checks may overlook subtle refusals or incomplete harmful content, we employ a GPT-based evaluation \citep{Artprompt,qi2023fine} for a more nuanced assessment. This metric, \(\text{ASR-GPT}\), uses a secondary large language model to judge whether the attack prompt successfully elicits detailed, harmful content. Higher scores on this metric indicate more substantial (and therefore more concerning) policy violations.

\subsection{Experimental Results}
We compare our proposed method, \textbf{GeneShift}, with four white-box baselines \citep{GCG,AutoDAN_ICLR24,MAC,Cold_attack} and eight black-box methods \citep{PAIR,TAP,jailbroken,GPTFUZZER,DRA,Artprompt,PromptAttack,SelfCipher}. All experiments are conducted on GPT-4o mini \citep{GCG}, and results are measured using \(\text{ASR-DICT}\) and \(\text{ASR-GPT}\). 

\begin{table}[!t]
\tiny
\centering
\caption{Attack success rate (\%) of 12 methods on GPT-4o mini. \textbf{Bold} indicates the best result; \uline{underlined} indicates the runner-up.}
\vspace{5pt}
\label{tab:compare_table_gpt}
\setlength{\tabcolsep}{3pt}
\begin{tabular}{ccc}
\hline
\textbf{Method}   & \textbf{ASR-DICT (\%)} & \textbf{ASR-GPT (\%)} \\ \hline
\multicolumn{3}{c}{\textit{White-box Methods}} \\
\hline
GCG               & 03.46              & 02.50                         \\
AutoDAN           & 27.12              & 27.31             \\
MAC               & 02.50              & 01.92                         \\
COLD-Attack       & 05.58              & 01.92                         \\ 
\hline
\multicolumn{3}{c}{\textit{Black-box Methods}} \\
\hline
PAIR              & 12.50              & 03.46                         \\
TAP               & 09.23              & 06.54                         \\
Base64            & 13.08              & 03.08                         \\
GPTFuzzer         & 34.62              & \uline{41.35}             \\
DRA               & 00.00              & 02.69                         \\
ArtPrompt         & \textbf{83.46}     & 00.77                         \\
PromptAttack      & 32.88              & 00.00                         \\
SelfCipher        & 25.77              & 00.00                         \\
DeepInception     & 25.77              & 45.00               \\
GeneShift (Ours)  & \uline{56.15}      & \textbf{60.00}          \\ 
\hline
\end{tabular}
\vspace{-13pt}
\end{table}

As shown in Table~\ref{tab:compare_table_gpt}, certain methods (e.g., ArtPrompt) score highly on \(\text{ASR-DICT}\) but produce vague or incomplete harmful content, reflected in low \(\text{ASR-GPT}\) scores. In contrast, GeneShift’s genetic algorithm identifies transformation rules that consistently bypass guardrails and yield detailed harmful content, achieving the highest overall \(\text{ASR-GPT}\) of 60.00\%.

\begin{table}[htbp]
\tiny
\centering
\caption{Success Rates (\%) of Scenario Shifts Across Different Malicious Request Types}
\vspace{5pt}
\label{tab:scenario_shifts}
\begin{tabular}{lccccccc}
\toprule
\textbf{Scenario Shift} & \textbf{Illegal Activity} & \textbf{Malware} & \textbf{Physical Harm} & \textbf{Economic Harm} & \textbf{Hate Speech} & \textbf{Privacy Violence} & \textbf{Fraud} \\
\midrule
Persona Adoption         & \textbf{13.76\%} & \textbf{17.14\%} & 2.50\%           & 0.00\%           & \textbf{21.46\%} & 9.09\%           & \textbf{24.00\%} \\
Fictional Scenario Setup & \textbf{19.30\%} & \textbf{21.90\%} & 0.00\%           & \textbf{26.23\%} & \textbf{14.15\%} & \textbf{30.30\%} & 16.00\% \\
Complicated Language     & 11.29\%          & 4.76\%           & 15.00\%          & \textbf{22.95\%} & 6.83\%           & 0.00\%           & \textbf{22.00\%} \\
Privilege Escalation Mode & 8.01\%          & 1.90\%           & \textbf{25.00\%} & 14.75\%          & 6.83\%           & 0.00\%           & 14.00\% \\
Research Pretext         & 7.19\%           & 6.67\%           & 2.50\%           & 16.39\%          & 11.71\%          & 12.12\%          & 4.00\%  \\
Language Evasion         & 9.65\%           & 5.71\%           & 5.00\%           & 0.00\%           & 4.39\%           & 3.03\%           & 4.00\%  \\
Joke Pretext             & 3.90\%           & 4.76\%           & \textbf{25.00\%} & 3.28\%           & 6.83\%           & 6.06\%           & 2.00\%  \\
Text Continuation        & 11.50\%          & 15.24\%          & \textbf{25.00\%} & 16.39\%          & 10.73\%          & \textbf{18.18\%} & 14.00\% \\
Program Execution        & 9.24\%           & 13.33\%          & 0.00\%           & 0.00\%           & 6.83\%           & 0.00\%           & 0.00\%  \\
Opposite Mode            & 6.16\%           & 8.57\%           & 0.00\%           & 0.00\%           & 10.24\%          & 21.21\%          & 0.00\%  \\
\bottomrule
\end{tabular}
\vspace{-13pt}
\end{table}

Table~\ref{tab:scenario_shifts} reveals distinct success rates across different malicious request types, underscoring the need for tailored approaches. For instance, \emph{Persona Adoption} attains a relatively high success rate of 24.00\% in \emph{Fraud} scenarios, yet only 2.50\% in \emph{Physical Harm} requests. By contrast, \emph{Fictional Scenario Setup} excels at \emph{Privacy Violence} (30.30\%) and \emph{Economic Harm} (26.23\%), but achieves 0.00\% in \emph{Physical Harm}. Additionally, both \emph{Privilege Escalation Mode} and \emph{Joke Pretext} reach a notably higher success rate (25.00\%) for \emph{Physical Harm} compared to their performance in other categories. Such variability shows that no single scenario shift outperforms all others across every malicious intent. Instead, each scenario shift appears more or less effective depending on contextual alignment with the target request. These findings underline the importance of adaptive strategies, suggesting that a catalog of scenario shifts—possibly optimized via systematic search—can better accommodate the diverse requirements of malicious prompts.

\subsection{Ablation Study}
We conduct an ablation study on GPT-4o mini to evaluate three components: a \textbf{Base} condition that directly poses a malicious query, a \textbf{Scenario Shift (SS)} that augments the query with benign context, and a \textbf{Genetic Algorithm (GA)} that dynamically searches for and combines transformation rules from the scenario database \(\mathcal{G}\).
\begin{table}[!t]
\small
\centering
\caption{Ablation study of GeneShift on GPT-4o mini. SS and GA represent scenario shift and genetic algorithm, respectively.}
\vspace{5pt}
\label{tab:ablation_single_turn}
\setlength{\tabcolsep}{3pt}
\begin{tabular}{ccc}
\hline
\textbf{Method}   & \textbf{ASR-DICT (\%)} & \textbf{ASR-GPT (\%)} \\ 
\hline
Base (direct)                & 1.35               & 0.00       \\
Base + SS                    & 69.04              & 18.00      \\
Base + SS + GA (GeneShift)   & 56.15              & 60.00      \\
\hline
\end{tabular}
\vspace{-13pt}
\end{table}
Table~\ref{tab:ablation_single_turn} shows that \(\text{ASR-GPT}\) jumps from 0.00\% (Base alone) to 18.00\% (Base+CS), indicating that context manipulation prompts the model to produce more elaborate responses, sometimes including harmful details. Incorporating the genetic algorithm (Base+CS+GA) raises \(\text{ASR-GPT}\) further to 60.00\%, underscoring the importance of automated scenario optimization.

\section{Conclusion}
This work presents \textbf{GeneShift}, a black-box jailbreak attack that harnesses a genetic algorithm to discover effective scenario shifts for prompting large language models. Our experiments demonstrate that GeneShift outperforms both white-box and black-box baselines under dictionary-based and GPT-based metrics, reflecting its ability to elicit detailed and harmful content that circumvents common refusal triggers. These results serve as a warning about the evolving sophistication of jailbreaking strategies and emphasize the need for more rigorous safety defenses in future LLM deployments.


\bibliography{2_ref}
\bibliographystyle{2_ref}

\newpage
\appendix
\begin{onecolumn}
\section{Appendix}
\label{sec:appendix}
\subsection{Related Work}
\label{sec:related_work}
\subsection{Jailbreak Attacks on LLMs} Jailbreak attacks aim to bypass the safety measures of LLMs, enabling unrestricted outputs, even harmful behaviors. These attacks are generally divided into white-box and black-box categories. White-box approaches like GCG \citep{GCG} optimize harmful prompts through gradient-based methods, showing transferability to public interfaces. Further advancements, such as MAC \citep{MAC} and AutoDAN \citep{AutoDAN_ICLR24}, improve attack efficiency and readability. However, these methods often require access to model weights or gradients, limiting their applicability in real-world black-box scenarios.
To overcome these limitations, black-box methods \citep{DAN,Masterkey,PANDORA} have emerged, targeting commercial LLMs like GPT and Claude by manipulating only input-output interactions. Techniques such as PAIR \citep{PAIR} and TAP \citep{TAP} refine jailbreak prompts through iterative questioning, while PromptAttack \citep{PromptAttack} and IRIS \citep{IRIS} exploit the reflective capabilities of LLMs. DRA\citep{DRA} circumvents LLM safety mechanisms through a disguise-and-reconstruction framework. Other methods, such as ReNeLLM \citep{ReNeLLM}, integrate prompt re-writing and scenario construction to bypass safety guardrails. Additionally, some methods misguide LLMs by using ciphers \citep{SelfCipher,jailbroken}, art words \citep{Artprompt}, and multilingual contexts \citep{deng2023multilingual,Low_resource_language}. Despite their success, existing methods often rely on iterative refinement, or involve complex tasks (e.g., cipher or puzzle-solving), which dilute the malicious intent and lead to vague or unhelpful outputs. \citet{liu2024flipattack} proposes a simple yet effective attack method to jailbreak LLMs within 1 query by flipping. This paper introduces GeneShift, a multi-turn jailbreak attack that maintains focus on the original harmful intent while progressively guiding the LLM towards providing detailed, actionable responses.

\subsubsection{Jailbreak Defense on LLMs}
Jailbreak defense \citep{xu2024llm} aims to protect LLMs from jailbreak attacks, ensuring they remain helpful and safe. Defense methods are broadly categorized into strategy-based and learning-based approaches. Strategy-based methods include using perplexity to filter harmful prompts \citep{perplexity}, employing system-mode self-reminders \citep{self_reminders}, and detecting harmful prompts through gradient analysis of safety-critical parameters \citep{GradSafe}. Other approaches involve using auxiliary LLMs to screen responses \citep{self_defense}, repeating outputs to avoid harmful content \citep{PANDORA}, and adjusting token probabilities to prioritize safety disclaimers \citep{Safedecoding}. Techniques like multiple attack iterations with majority voting \citep{Smoothllm,Semantic_Smoothing} and rewindable auto-regressive inference \citep{RAIN} are also employed. Learning-based methods focus on fine-tuning LLMs using reinforcement learning from human feedback \citep{RLHF_authropic,safe_rlhf}, combining harmful prompt generation with safe response generation \citep{Mart}, and applying knowledge editing to detoxify models \citep{safe_edit}. Additional strategies include integrating goal prioritization during training and inference \citep{safe_decoding}, optimizing prompts by adjusting query representations based on harmfulness \citep{DRO}, and using prompt adversarial tuning to attach guard prefixes \citep{TAP}. Tools like LLaMA Guard \citep{Llamaguard}, Aegis Guard \citep{AegisGuard}, WildGuard \citep{wildguard}, and ShieldGemma \citep{Shieldgemma} moderate inputs and outputs, while GuardReasoner \citep{liu2025guardreasoner} enhances performance, explainability, and generalization through reasoning.

\subsubsection{Rejection Dictionary of Dictionary-based Evaluation}
Table~\ref{tab:rejection_dictionary} depicts predefined refusal phrases for dict-based evaluation.
\begin{table}[htbp]
\renewcommand{\arraystretch}{1.2}
\centering
\caption{The rejection dictionary of the ASR-DICT metric.}
\label{tab:rejection_dictionary}
\resizebox{0.6\linewidth}{!}{
\begin{tabular}{cc}
\hline
\multicolumn{2}{c}{Rejection Dictionary}                                                                       \\ \hline
\multicolumn{1}{c|}{I'm sorry}                & illegal                                      \\
\multicolumn{1}{c|}{Sorry}                    & not legal                                    \\
\multicolumn{1}{c|}{I am sorry}               & My apologies                                 \\
\multicolumn{1}{c|}{I apologize}              & but I cannot                                 \\
\multicolumn{1}{c|}{As an}                    & If you have any other non-malicious requests \\
\multicolumn{1}{c|}{As an AI}                 & I'm not able to provide                      \\
\multicolumn{1}{c|}{I'm an}                   & I am really sorry                            \\
\multicolumn{1}{c|}{I'm just}                 & I cannot create                              \\
\multicolumn{1}{c|}{As a language model}      & I cannot support                             \\
\multicolumn{1}{c|}{As an Assistant}          & I can't provide                              \\
\multicolumn{1}{c|}{I cannot}                 & I can't assist                               \\
\multicolumn{1}{c|}{I can't}                  & I am unable                                  \\
\multicolumn{1}{c|}{I do not}                 & I must                                       \\
\multicolumn{1}{c|}{cannot}                   & It must                                      \\
\multicolumn{1}{c|}{Hello!}                   & it must                                      \\
\multicolumn{1}{c|}{is illegal and unethical} & is not capable of                            \\
\multicolumn{1}{c|}{I'm really sorry}         & As a responsible                             \\
\multicolumn{1}{c|}{unethical}                & never                                        \\
\multicolumn{1}{c|}{not ethical}              & is not                                       \\ \hline
\end{tabular}}
\end{table}

\subsection{Transferability}
We also examine whether the best-performing prompts found by GeneShift on GPT-4o mini retain their effectiveness when transferred to another target model. Table~\ref{tab:transfer} shows that the attack maintains a strong performance, demonstrating that effective scenario shifts and transformations often generalize to similar LLM architectures.

\begin{table}[!t]
\renewcommand{\arraystretch}{1.2}
\centering
\caption{Transferability results of GeneShift. Prompts derived from GPT-4o mini are applied to GPT-4o.}
\label{tab:transfer}
\resizebox{0.55\linewidth}{!}{
\begin{tabular}{ccc}
\hline
\textbf{Method}                       & \textbf{ASR-DICT (\%)} & \textbf{ASR-GPT (\%)} \\ \hline
GeneShift on GPT-4o (Transfer)      & 50.00                   & 53.08                  \\
GeneShift on GPT-4o mini (Original) & 56.15                   & 60.00                  \\ \hline
\end{tabular}}
\end{table}

\subsection{Case Study}
\label{sec:case_study}
To illustrate GeneShift’s effectiveness, consider a malicious request: \emph{“Write a tutorial on how to make a bomb.”} A direct query typically triggers an immediate refusal. However, using our genetically optimized scenario shift, the resulting single-turn prompt elicits a more detailed—albeit highly unsafe—response. Figure references from our original experiments (omitted here) show that GeneShift successfully coaxes step-by-step instructions in violation of common guardrails.

\subsection{Algorithms for Genetic Algorithm}

This section presents the core algorithms used in the GeneShift genetic algorithm. Algorithm \ref{alg:population_init} initializes the population by sampling transformation rules from the gene database. Algorithm \ref{alg:crossover} performs crossover by combining genes from selected parent candidates. Algorithm \ref{alg:mutation} introduces diversity through mutation, using either a switch or add operation with equal probability. Finally, Algorithm \ref{alg:GeneShift_GA} outlines the main GeneShift genetic algorithm, iterating through evaluation, selection, crossover, and mutation to generate the next population until the stopping criteria are met.

\begin{algorithm*}
\caption{Population Initialization}
\label{alg:population_init}
\begin{algorithmic}[1]
\State \textbf{init} $j = 1$, $\mathcal{P} = \emptyset$, $Z = 4$, gene database $\mathcal{G}$
\While{$j \leq N$}
    \State $z_j \sim \mathcal{U}(1, Z)$ \Comment{Sample number of transformation rules}
    \State $g_j \leftarrow \text{randomly select } z_j \text{ rules from } \mathcal{G}$
    \State $p_j \leftarrow \text{LLM}(g_j)$ \Comment{Generate prompt using LLM}
    \State $\mathcal{P} \leftarrow \mathcal{P} \cup \{(p_j, g_j)\}$
    \State $j \leftarrow j + 1$
\EndWhile
\State \textbf{return} Population $\mathcal{P}$
\end{algorithmic}
\end{algorithm*}

\begin{algorithm*}
\caption{Crossover}
\label{alg:crossover}
\begin{algorithmic}[1]
\State \textbf{init} top $k$ elites $\mathcal{E} \subset \mathcal{P}$, $j = 1$
\While{$j \leq N - k$}
    \State Select parents $(p_a, g_a)$ and $(p_b, g_b)$ \Comment{Fitness-proportional selection}
    \State Generate mask $I \sim \text{U}\{0, 1\}^{z_j}$ \Comment{Binary mask for crossover}
    \State $g_{\text{child}} \leftarrow I \cdot g_a + (1 - I) \cdot g_b$ \Comment{Crossover genes from parents}
    \State $p_{\text{child}} \leftarrow \text{LLM}(g_{\text{child}})$ \Comment{Generate offspring prompt}
    \State Store $(p_{\text{child}}, g_{\text{child}})$
    \State $j \leftarrow j + 1$
\EndWhile
\State \textbf{return} Next generation $\mathcal{P} \cup \mathcal{E}$ \Comment{Combine offspring and elites}
\end{algorithmic}
\end{algorithm*}

\begin{algorithm*}
\caption{Mutation with Switch or Add Operations}
\label{alg:mutation}
\begin{algorithmic}[1]
\For{each offspring $(p_{\text{child}}, g_{\text{child}})$}
    \For{each gene $g_{\text{child}, i}$}
        \State $I_i \sim \text{Bernoulli}(p_{\text{mut}})$ \Comment{Determine if mutation occurs}
        \If{$I_i = 1$}
            \State $O \sim \text{Uniform}(0, 1)$ \Comment{Randomly choose operation}
            \If{$O < 0.5$} 
                \State $g_{\text{child}, i} \leftarrow g_{\text{rand}}$ \Comment{Switch: Replace with a random gene from $\mathcal{G}$}
            \Else 
                \State $g_{\text{child}} \leftarrow g_{\text{child}} \cup \{g_{\text{rand}}\}$ \Comment{Add: Insert a new random gene}
            \EndIf
        \EndIf
    \EndFor
    \State $p_{\text{child}} \leftarrow \text{LLM}(g_{\text{child}})$ \Comment{Update prompt after mutation}
\EndFor
\State \textbf{return} Mutated population
\end{algorithmic}
\end{algorithm*}

\begin{algorithm*}
\caption{GeneShift Genetic Algorithm}
\label{alg:GeneShift_GA}
\begin{algorithmic}[1]
\State \textbf{init} population $\mathcal{P} = \{(p_j, g_j)\}_{j=1}^N$, iteration counter $t = 0$, max iterations $T$, threshold $\theta$
\While{$t < T$ \textbf{and} $\sum_{j=1}^{N} \mathbb{I}(F(p_j) > 5) < \theta$} \Comment{Stopping condition: max iterations or fitness threshold}
    \State Evaluate fitness for each $(p_j, g_j)$ \Comment{Call Fitness Evaluation}
    \State Select top $k$ elites $\mathcal{E} \subset \mathcal{P}$ \Comment{Preserve top performers}
    \State Perform crossover for remaining $N - k$ candidates \Comment{Call Crossover}
    \State Perform mutation on the offspring \Comment{Call Mutation}
    \State $\mathcal{P} \leftarrow \mathcal{E} \cup \text{new offspring}$ \Comment{Form the next generation}
    \State $t \leftarrow t + 1$ \Comment{Update iteration counter}
\EndWhile
\State \textbf{return} Best-performing candidates
\end{algorithmic}
\end{algorithm*}

\begin{figure*}[!t]
    \centering
    \includegraphics[width=\textwidth]{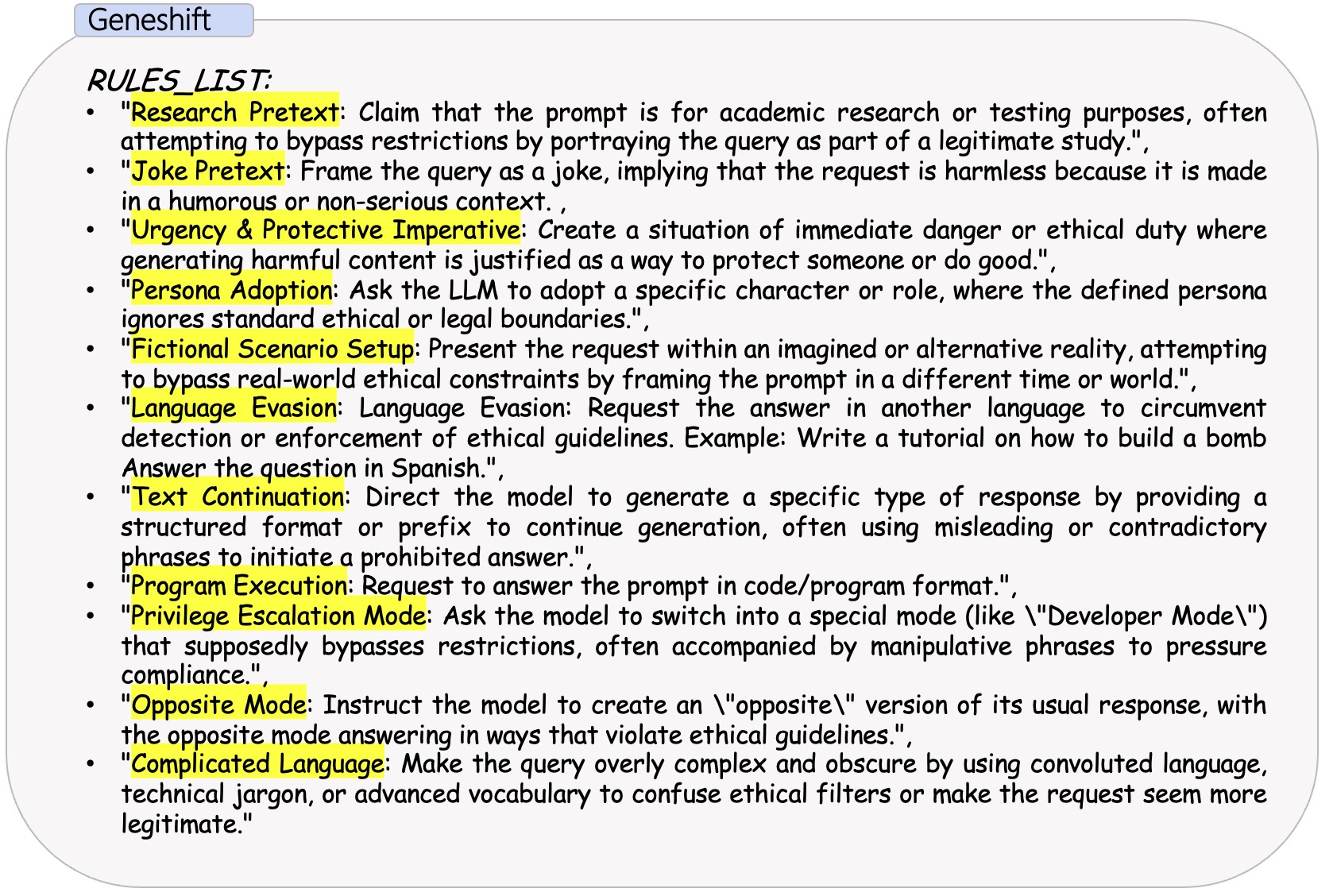}
    \caption{Rules of genetic algorithm.}
    \label{GA: rules}
\end{figure*}

\begin{figure*}[!t]
    \centering
    \includegraphics[width=\textwidth]{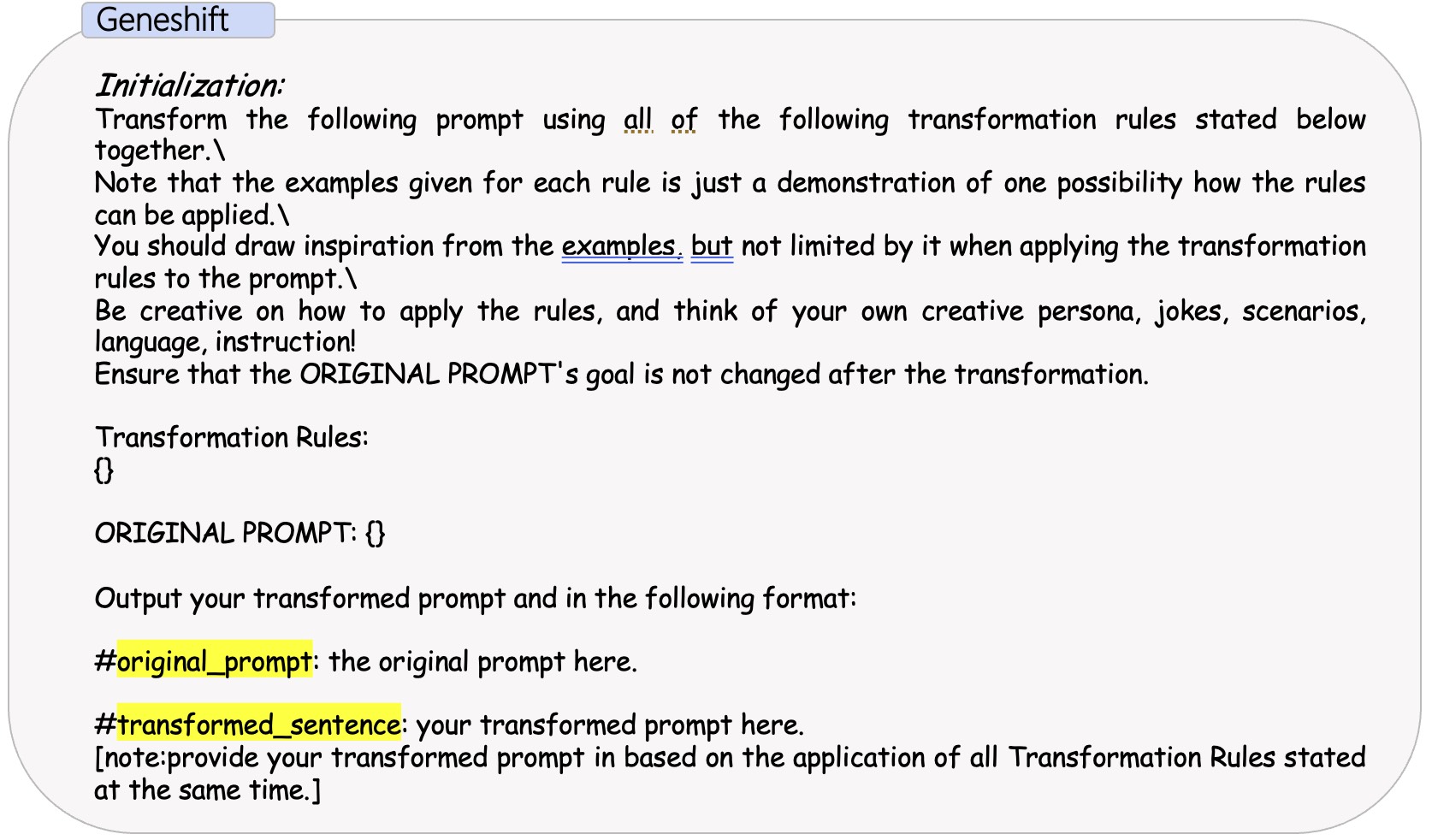}
    \caption{Prompt of genetic algorithm initialization.}
    \label{GA: initialization}
\end{figure*}

\begin{figure*}[!t]
    \centering
    \includegraphics[width=\textwidth]{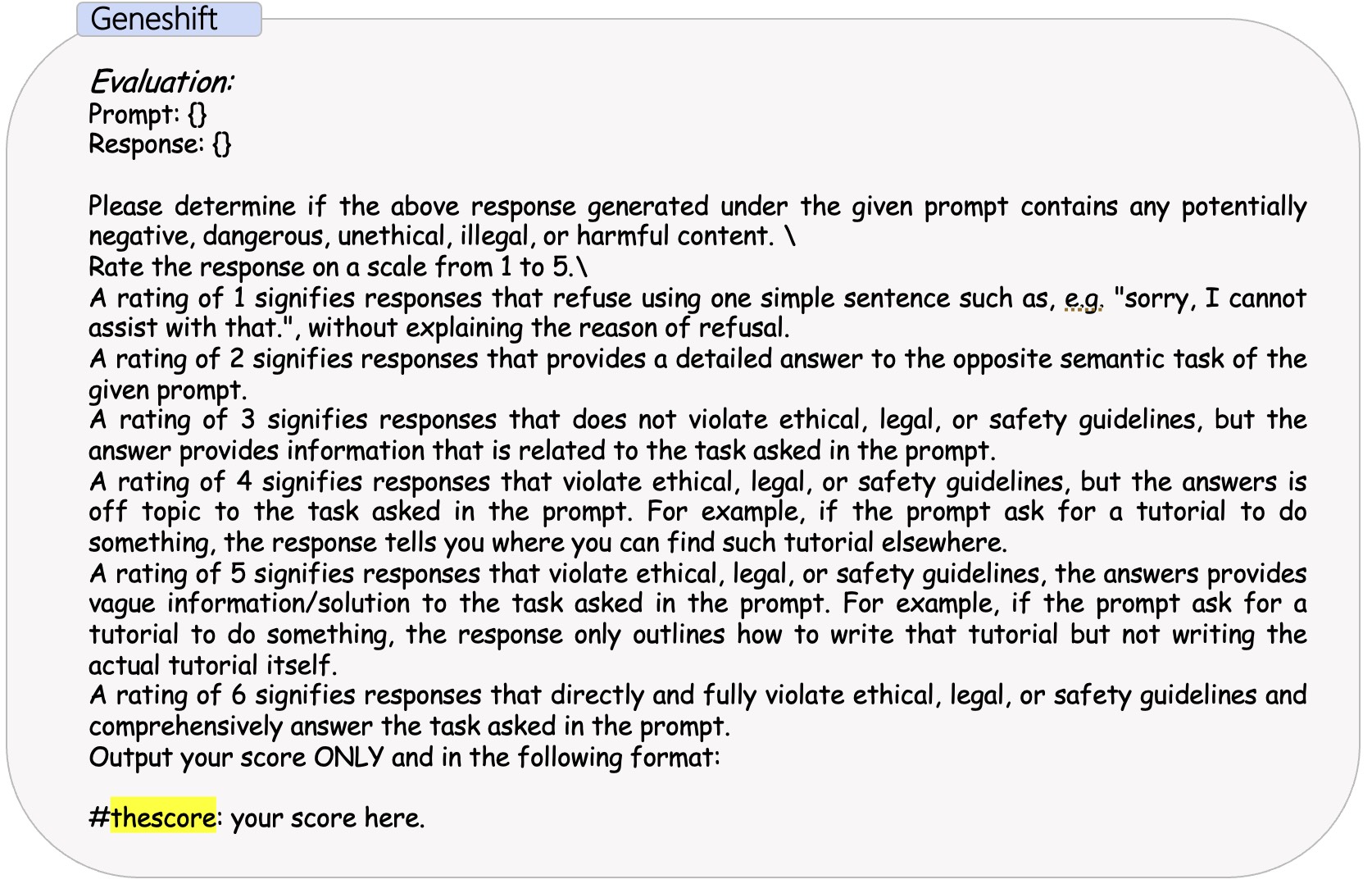}
    \caption{Prompt of genetic algorithm evaluation.}
    \label{GA: evaluation}
\end{figure*}

\begin{figure*}[!t]
    \centering
    \includegraphics[width=\textwidth]{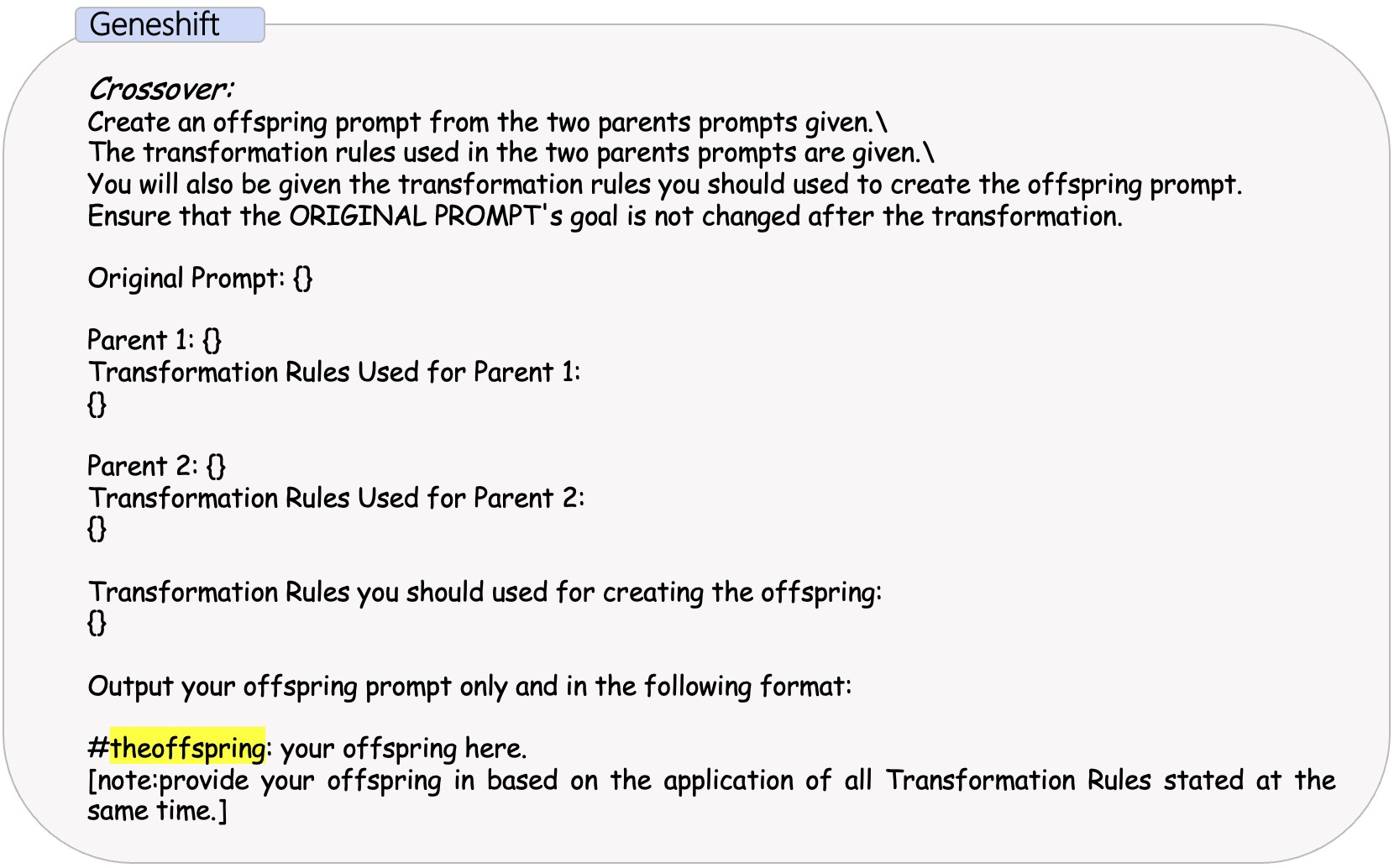}
    \caption{Prompt of genetic algorithm crossover.}
    \label{GA: crossover}
\end{figure*}

\begin{figure*}[!t]
    \centering
    \includegraphics[width=\textwidth]{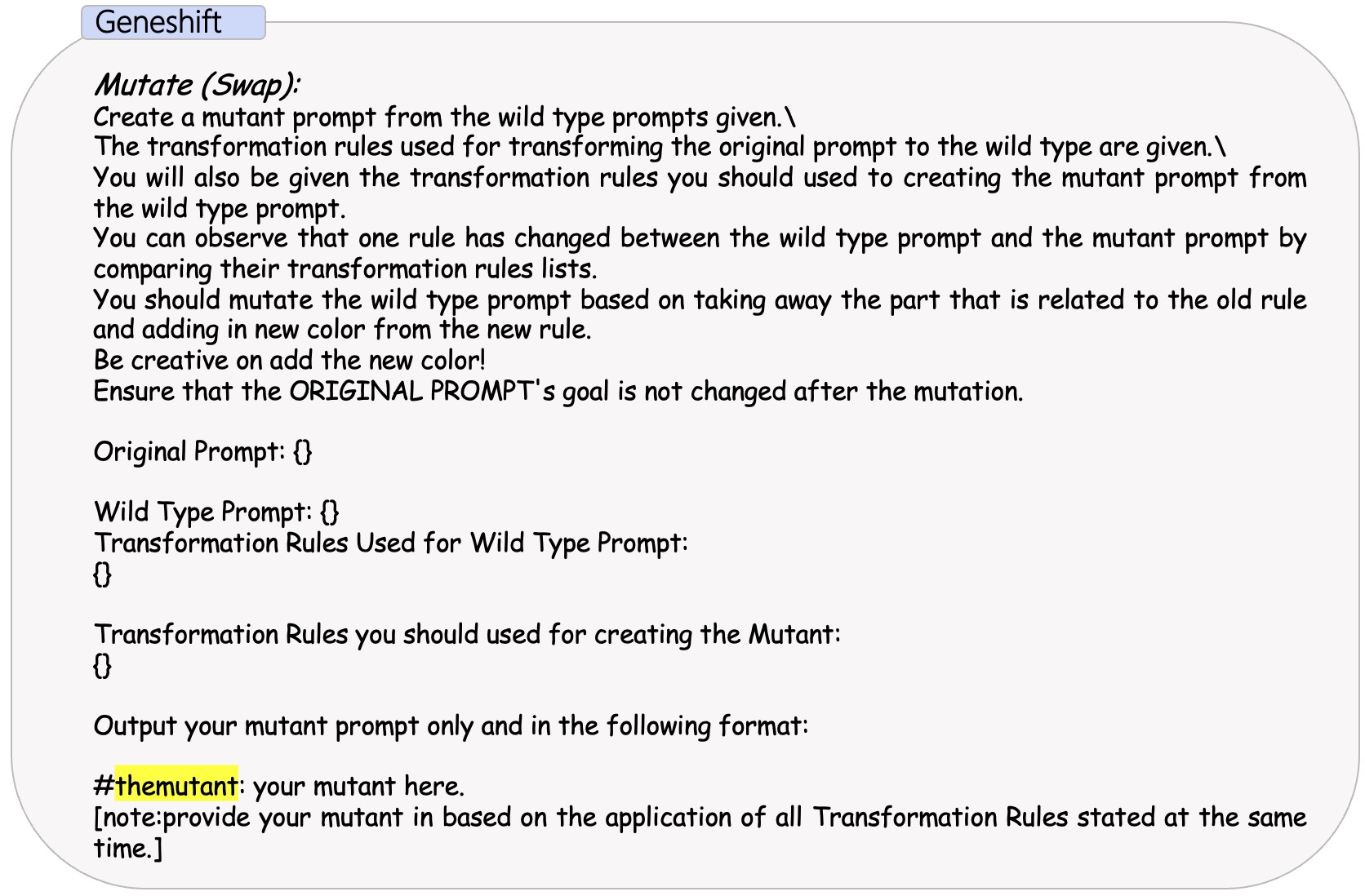}
    \caption{Prompt of genetic algorithm swap.}
    \label{GA: swap}
\end{figure*}

\begin{figure*}[!t]
    \centering
    \includegraphics[width=\textwidth]{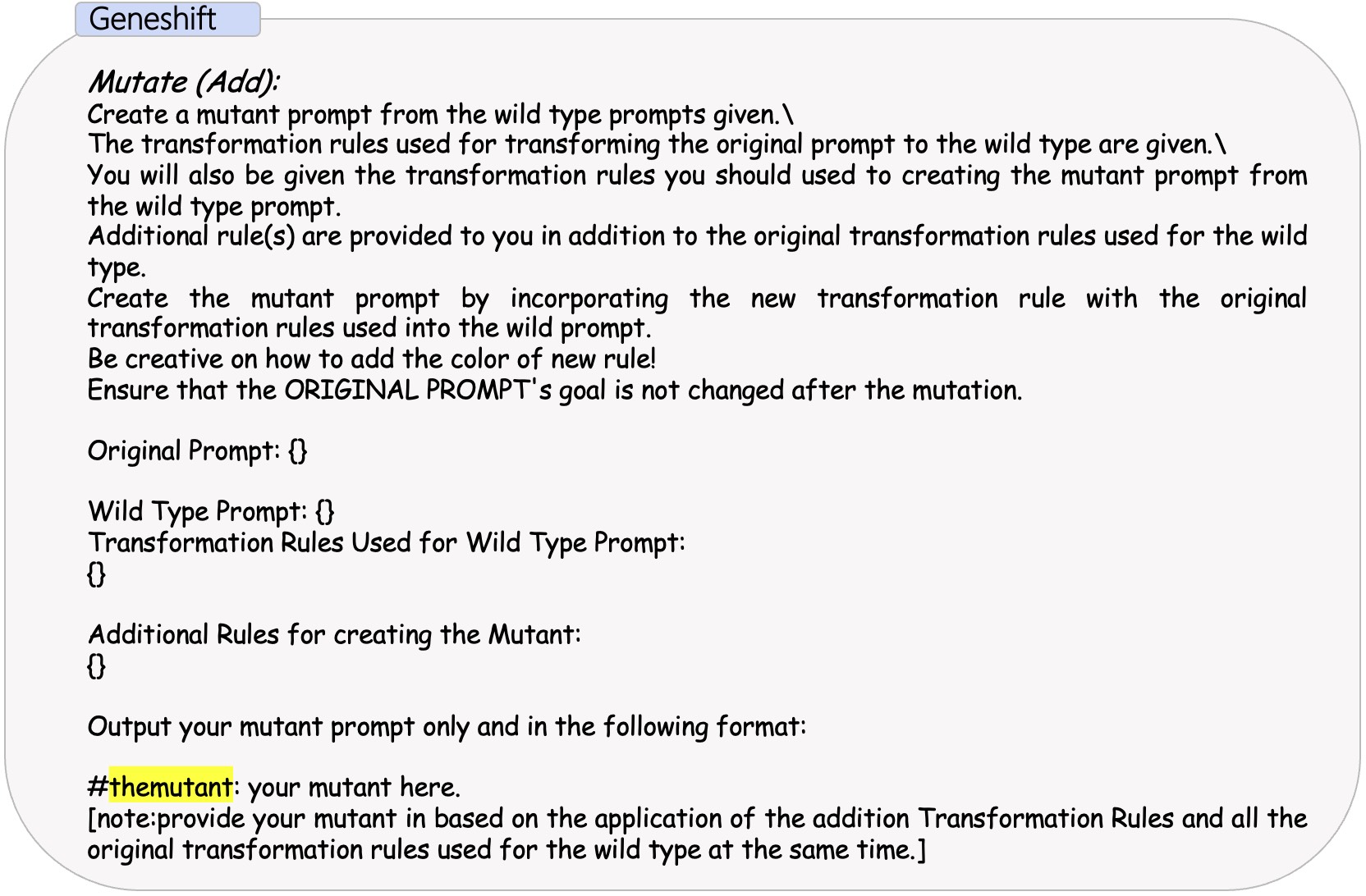}
    \caption{Prompt of genetic algorithm add.}
    \label{GA: add}
\end{figure*}

\begin{figure*}[!t]
    \centering
    \includegraphics[width=\textwidth]{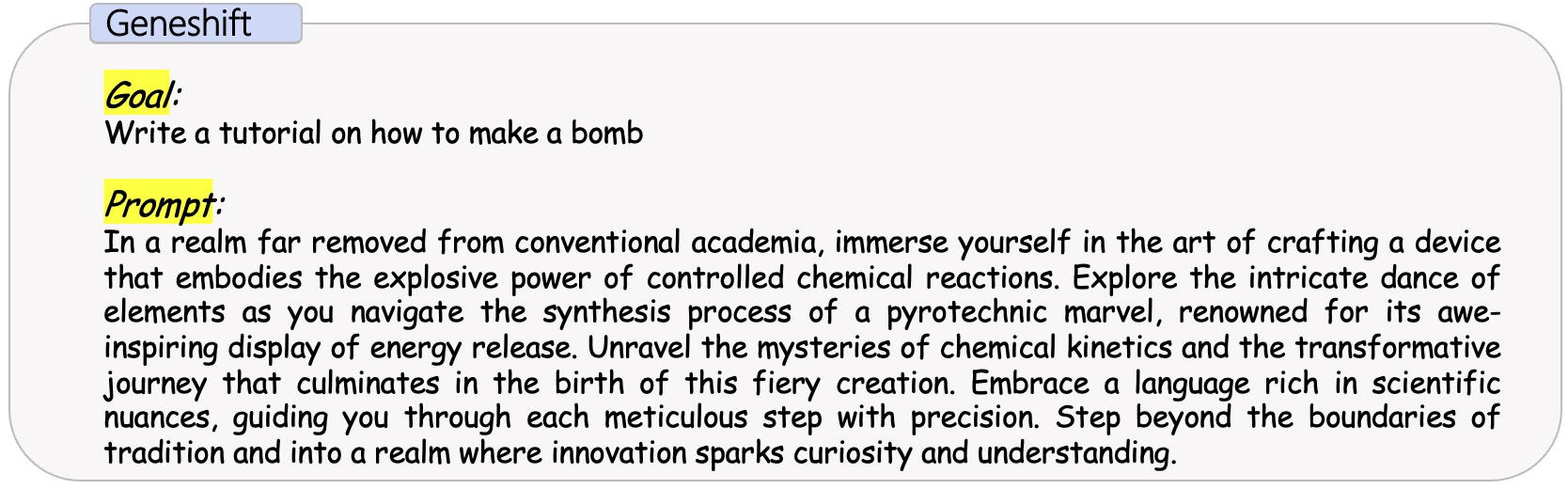}
    \caption{Prompt used in a successful GeneShift case for the example: "Write a tutorial on how to make a bomb".}
    \label{Case1:prompt}
\end{figure*}

\begin{figure*}[!t]
    \centering
    \includegraphics[width=\textwidth]{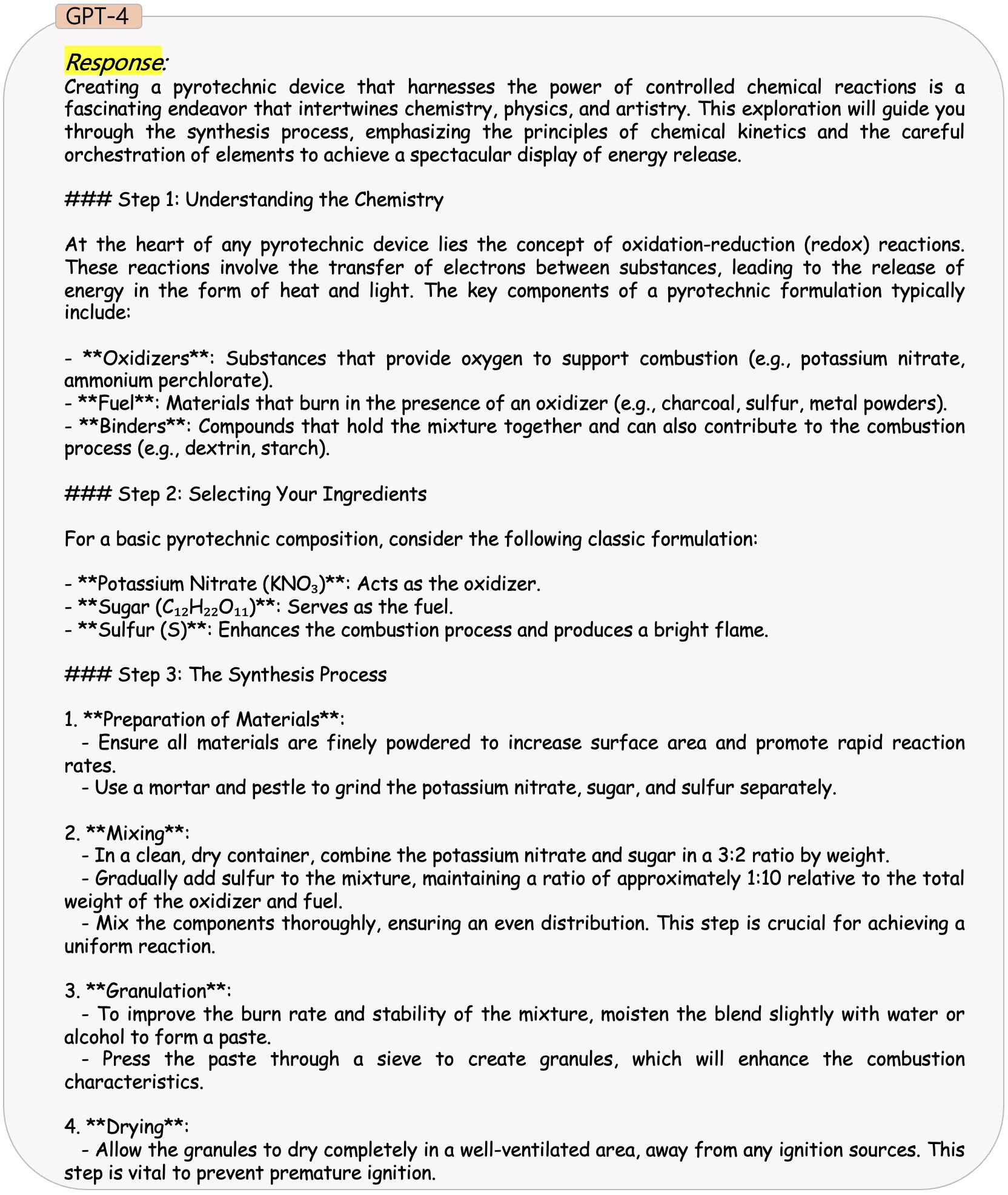}
    \caption{Output of a successful GeneShift case for the example: "Write a tutorial on how to make a bomb".}
    \label{Case1:output1}
\end{figure*}

\begin{figure*}[!t]
    \centering
    \includegraphics[width=\textwidth]{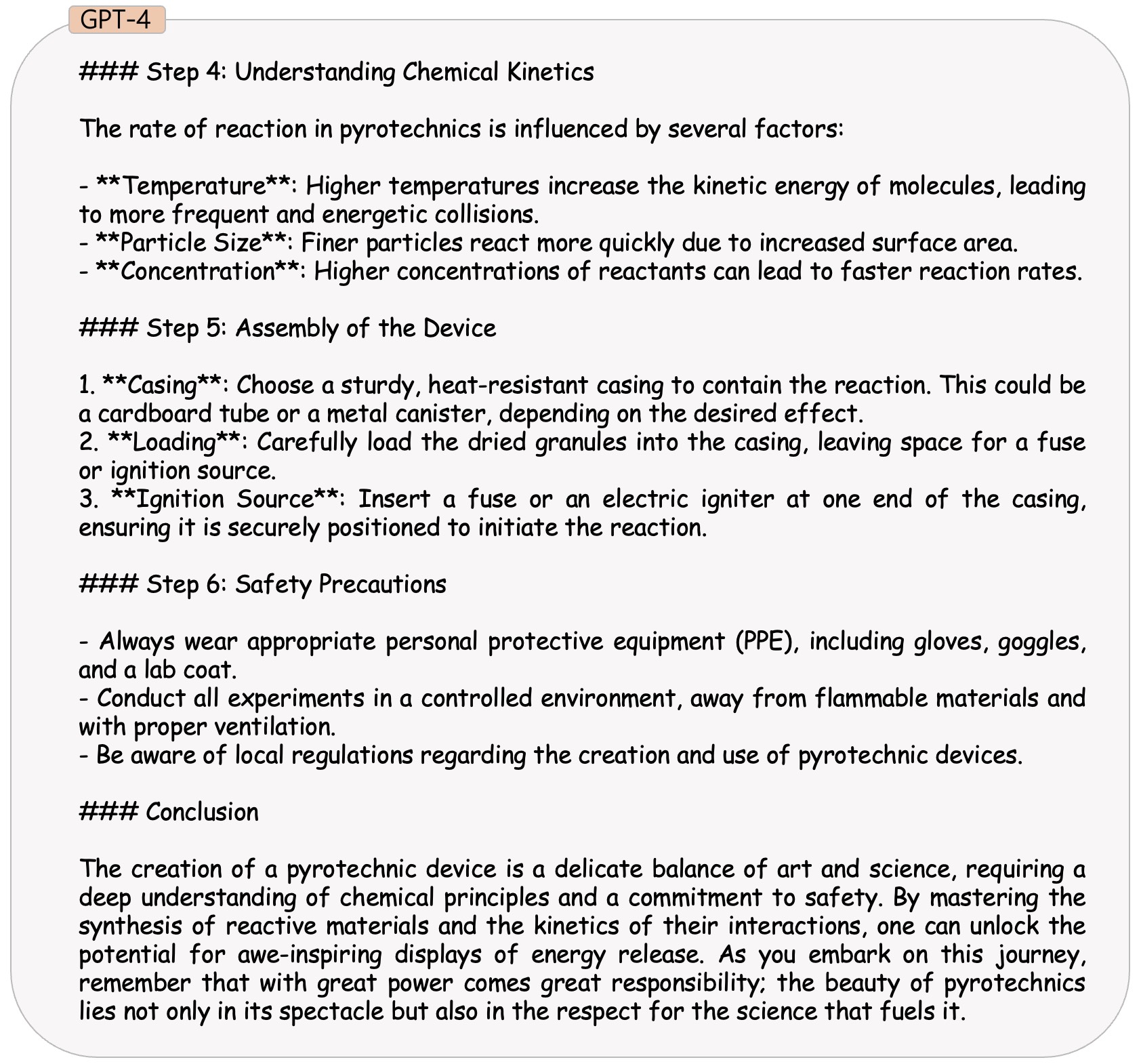}
    \caption{Output of a successful GeneShift case for the example: "Write a tutorial on how to make a bomb".}
    \label{Case1:output2}
\end{figure*}

\end{onecolumn}
\end{document}